\documentclass[aps,pre,preprint,groupedaddress,showpacs]{revtex4-1}
\usepackage[dvips]{graphicx}
\usepackage{textcomp}
\usepackage{amssymb}

\newcommand{\mc}{\multicolumn}

\begin{document}

\title{
\Large\bf Precision estimates of large charge RG exponents $Y_q$ in the 
3D XY universality class}

\author{Martin Hasenbusch}
\affiliation{
Institut f\"ur Theoretische Physik, Universit\"at Heidelberg,
Philosophenweg 19, 69120 Heidelberg, Germany}

\date{\today}
\begin{abstract}
We accurately compute the RG exponents $Y_q$ of large $q$ fields at 
the $O(2)$ invariant fixed point in three dimensions. We build 
on an iterative approach that has been previously proposed and is 
implemented by using the worm algorithm. We simulate an improved 
XY model, that has next-to-next-to-nearest couplings in addition 
to nearest ones. In the worm update we incorporate weights, which
allows us to obtain accurate results up to $q=64$. For example 
we get $Y_q=1.76370(12)$, $0.89167(23)$, and $-0.11203(34)$ 
for $q=2$, $3$, and $4$, respectively.
The comparison with the large $q$ effective field theory 
gives an excellent agreement down to $q=4$ and provides accurate 
estimates of the parameters of the effective field theory.
\end{abstract}

\maketitle

\section{Introduction}
Critical phenomena \cite{WiKo,Fisher74,Cardy,Fisher98,PeVi02} 
in three dimensions have been studied by 
using various theoretical approaches such as field theoretic methods, 
series expansions and Monte Carlo (MC) simulations of lattice models and 
more recently the conformal bootstrap (CB) method  \cite{PaRyVi18,RySu23}. 
The theoretical framework 
is the renormalization group (RG), which is the basis of various methods,
and provides us with fundamental notions such as universality.
It has been realized that in the large spin and large charge limit,
semi-classical approximations are viable. See for example Refs.
\cite{Alday,Fitz,Koma,Hellerman2015,Monin17}.
For a review see  \cite{Alvarez21}.
Here we are concerned with large charge $q$ in the simplest case, the 
XY-universality class in three dimensions.
The dimension  $D_q$ of the lowest operator with global charge 
$q$ behaves as \cite{Hellerman2015,Monin17,Alvarez21} 
\begin{equation}
\label{CentralEQ}
D_q = c_{3/2} q^{3/2} + c_{1/2} q^{1/2}  - 0.0937256 + O(q^{-1/2}) \;.
\end{equation}
Here we denote the dimension of the operator by $D_q$,
since later we shall use $\Delta_q=D_q-D_{q-1}$. The corresponding
RG exponent is $Y_q=d-D_q$, where here $d=3$ is the spatial dimension of
the system. The coefficients $c_{3/2}$ and $c_{1/2}$ are unknown, 
while the constant is predicted analytically and should not depend
on the universality class. The coefficients $c_{3/2}$ and $c_{1/2}$ are
universal, but differ between different universality classes.

The dimensions $D_q$ for $q \le 6$ have been computed directly 
by using various methods, in particular Monte Carlo simulations of 
lattice models, field theoretic methods, and the conformal bootstrap.
Irrespective of the method, the error rapidly increases with increasing
$q$. For a summary and a discussion of these results see Sec. \ref{summary}.

In Ref. \cite{Banerjee18} an approach is suggested that allows to 
reach larger values of $q$ by using Monte Carlo simulations of lattice 
models. The authors simulate the standard XY model on the simple cubic
lattice by using the worm algorithm \cite{Prok98,Prok01}.  
They study the two-point function at criticality:
\begin{equation}
\label{twopoint}
C_q(r) = \langle \exp(-i q (\theta_0 - \theta_r) ) \rangle
\simeq |r|^{-2 D_q} \;,
\end{equation}
where $x$ is a site on the finite lattice with the linear extension $L$.
In order to cancel finite size effects, they consider $r=(L/2,0,0)$ 
throughout. In a direct approach the error of $D_q$,
for a given numerical effort, would increase rapidly with $q$. The way out
is to compute ratios
\begin{equation}
\label{Gratio}
R_q(r) = \frac{C_q(r)}{C_{q-1}(r)} \simeq |r|^{-2 \Delta_q} \;,
\end{equation}
where $\Delta_q = D_q-D_{q-1}$. A discussion of the implementation is given 
below in Sec. \ref{worm3N}. 
The authors report results up to $q=12$, see table II of 
Ref. \cite{Banerjee18}. More recently the idea was picked up 
in Ref. \cite{Cuomo23} who added a new algorithmic idea and studied
three-point functions in addition.  The authors of \cite{Cuomo23}
reach $q=19$, albeit with a rather big error of $\Delta_q$. The numerical
results are in good agreement with Eq.~(\ref{CentralEQ})  down to 
astonishingly small values of $q$. 

Here we build on Ref. \cite{Banerjee18} and use the algorithmic idea of
Ref. \cite{Cuomo23}. However, instead of the standard XY model
on the simple cubic lattice, we simulate the improved model studied 
in Ref. \cite{myXY2025}. Improved means that the amplitude of the leading 
correction to scaling vanishes. Major progress is achieved by
using non-trivial weights in the worm update  \cite{Ulli_Demo_09}. 
This allows us to reach $q=64$ at high numerical precision.
Our results show that both Refs. \cite{Banerjee18,Cuomo23} have underestimated
systematic errors due to corrections to scaling. Our errors on 
$\Delta_q$ are in the fourth digit. The results are in excellent agreement 
with Eq.~(\ref{CentralEQ}) down to $q \approx 4$. The deviation is small even
for $q=1$. This certainly calls for a better theoretical understanding.

The outline of the paper is the following: First, in Sec. \ref{themodel}
we define the 3N XY model that we simulate and the quantities that we measure.
Then in Sec. \ref{worm3N} we explain how the 3N XY model is simulated by 
using the worm algorithm and how the ratio of two-point functions is 
implemented. In Sec. \ref{Numerics} we discuss the simulations and the 
analysis of the resulting data. Sec. \ref{Numerics} starts with its own
outline. Finally we summarize our results and compare with the literature.
In the appendix we provide some additional discussion on the worm algorithm.

\section{The model and the two-point function}
\label{themodel}
We consider the XY model on the simple cubic lattice with 
nearest neighbor and next-to-next-to-nearest neighbor (3N) couplings.
Below we shall denote the model by 3N XY model.
The reduced Hamiltonian is given by
\begin{equation}
\label{Hamilton}
 {\cal H}[\{\vec{s}\}] = -  K_1 \sum_{\left<xy\right>}  \vec{s}_x \cdot
     \vec{s}_y 
  -  K_3 \sum_{[xy]} \vec{s}_x \cdot \vec{s}_y \;,
\end{equation}
where the spin or field variable $\vec{s}_x$ is a unit vector with two 
real components. It can be written as 
$\vec{s}_x=[\cos(\theta_x), \sin(\theta_x)]$, where $\theta_x \in [0, 2 \pi)$.
$x=(x_0,x_1,x_2)$ denotes a site on the simple cubic lattice,
where $x_i \in \{0,1,...,L_i-1\}$. Furthermore, $<xy>$ denotes a pair
of nearest and $[xy]$ a pair of next-to-next-to-nearest, or third nearest (3N)
neighbors on the lattice. In our simulations,
the linear lattice size $L=L_0=L_1=L_2$ is equal in all three directions
and periodic boundary conditions are employed.
The partition function is given by
\begin{equation}
 Z = \int \mbox D[\{\vec{s}\}] \exp(- {\cal H}[\{s\}]) \;.
\end{equation}
In Ref. \cite{myXY2025} we have set up the FSS study such that
we pass the critical line for a given value of $K_3$ by varying $K_1$.  
Hence we get a critical coupling $K_{1,c}$ that depends on $K_3$. 
The standard XY model is obtained for $K_3=0$. 

We find that leading corrections to scaling at criticality are eliminated at 
$K_3^*=0.04149(14)$.  For $K_3=0.0415$ we get $K_{1,c}= 0.37069947(2)$.
Below, we shall simulate at $(K_1,K_3)=(0.37069947,0.0415)$. 
Note that in Ref. \cite{myXY2025} we have actually simulated the $q$-state
clock model. For the values of $q$ that we have simulated, the difference 
in $K_{1,c}$ compared with the limit $q \rightarrow \infty$ is 
negligible.

The choice of the improved model is not unique. In the literature, for example,
the two-component $\phi^4$ model and the dynamically diluted XY model
are studied. In Refs. \cite{HaTo99,XY1,XY2,myClock,myXY2025} the values of 
the parameters, where leading corrections to scaling vanish are  
accurately determined.

Here we have selected the 3N XY model, since it is straight forward 
to use the worm algorithm as it has been implemented for the standard 
XY model in the literature. Furthermore in Ref. \cite{myXY2025} 
we have determined  $K_{1,c}$
and $K_3^*$ very accurately and corrections due to the violation of the 
rotational invariance are reduced compared with the standard XY model.

There are different ways to extract the dimensions of fields in Monte
Carlo simulations of lattice models. Here we follow the straight forward 
approach that has been used in Refs. \cite{Banerjee18,Cuomo23}.
At criticality the two-point function, Eq.~(\ref{twopoint}), behaves as
$C_q(r) \simeq |r|^{-2 D_q}$, where $D_q$ is the dimension of the field.
For a finite lattice this behavior is observed only 
for $|r| \ll L$, where $L$ is the linear lattice size. In practice, 
this requirement impedes precision estimates of $D_q$.  Therefore, 
following Ref. \cite{Banerjee18}, we chose $|r|= c L$ to cancel
finite size effects. This means, for a given linear lattice size $L$, only one 
distance is considered.  In the case of Ref. \cite{Banerjee18} $r =(L/2,0,0)$.
Here we decided ad hoc to use the maximal distance on the torus
\begin{equation}
\label{our_distance}
 r =(L/2,L/2,L/2)  \;.
\end{equation}
We did not carefully compare these two choices. Likely, the difference in 
statistical errors and systematic errors due to corrections to scaling is 
rather small.
The two-point function can be determined by using the worm algorithm as
we outline below.

\section{The worm algorithm for the XY model with 3N couplings}
\label{worm3N}
Here we follow previous work on the standard XY model, for example Refs. 
\cite{Banerjee10,Banerjee18,Deng19}. Taking care of the additional
3N coupling is straight forward.
In order to discuss the worm algorithm, the partition function 
has to be rewritten. A new field $\{k\}$ that lives on the edges
of the lattice is introduced, which allows to integrate out the 
original field $\{\vec{s}\}$ that lives on the sites. In the case of 
our model we have to add variables for the 3N neighbor pairs.
First we express the reduced Hamiltonian in terms of angles
\begin{equation}
H(\{\theta\})  = 
  -  K_1 \sum_{\left<xy\right>}  \cos(\theta_x -\theta_{y})
  -  K_3 \sum_{[xy]} \cos(\theta_x -\theta_{y}) \;,
\end{equation}
where we change variables as $\vec{s}_x = [\cos(\theta_x), \sin(\theta_x)]$ 
with $\theta_x \in [0,2 \pi)$. It holds
\begin{equation}
\exp(K \cos(\theta)) = \sum_{k=-\infty}^{\infty} I_k(K) 
\exp(i k \theta) \;,
\end{equation}
where $I_k(K)$ is the modified Bessel function of the first kind.
Hence the partition function becomes
\begin{eqnarray}
\label{XYpartition}
Z &=& \int \mbox{D}[\theta] \left[ \prod_{<xy>} \exp(K_1 \cos(\theta_x-\theta_y)) \right]
                           \left[ \prod_{[xy]} \exp(K_3 \cos(\theta_x-\theta_y)) \right] \nonumber \\
  &=& \int \mbox{D}[\theta] \sum_{\{k\}} \left[ \prod_{<xy>} I_{k_{<xy>}}(K_1)
        \exp(i k_{<xy>} [\theta_x-\theta_y]) \right] 
       \left[ \prod_{[xy]}  I_{k_{[xy]}}(K_3) \exp(i k_{[xy]} [\theta_x-\theta_y]) \right] \nonumber\\
&=&  \sum_{\{k\}} \left[ \prod_{<xy>} I_{k_{<xy>}}(K_1) 
          \int \mbox{D}[\theta] \exp(i k_{<xy>} [\theta_x-\theta_y]) \right] 
\nonumber\\
  && \phantom{xxxxxxxx}         \left[ \prod_{[xy]} I_{k_{[xy]}}(K_3) 
          \int \mbox{D}[\theta] \exp(i k_{[xy]} [\theta_x-\theta_y]) \right] \;,
\end{eqnarray}
where ${\{k\}}$ is a configuration of the $k$ variables.
We need a convention, for the orientation of the edges $<xy>$ and 
$[xy]$. Depending on the relative orientation, going from $x$ to $y$, 
$-1$ or $1$ is multiplied on $k_{<xy>}$ or $k_{[xy]}$. Hence 
$k_{<xy>}=-k_{<yx>}$ and $k_{[xy]}=-k_{[yx]}$.
In the case of $<xy>$ we assign the positive orientation to the
upward link. In the case of $[xy]$, the positive orientation goes to
$x_0$ up and the negative one to $x_0$ down. Of course, equivalently
we could take $x_1$ or $x_2$ to this end.
Performing the integral over $\theta$, we arrive at the constraint
\begin{equation}
\label{const0}
\sum_{y.nn.x}  k_{<xy>} + 
\sum_{y.3n.x}  k_{[xy]} =0
\end{equation}
for all $x$. Here $y.nn.x$ means that $y$ is a nearest neighbor of $x$ and
$y.3n.x$ that $y$ is a third nearest neighbor of $x$. 
The constraint is solved by packing closed oriented loops on
the lattice. For our model, the loop might run through the edges $<xy>$ as 
for the standard XY model with $K_3=0$, but now in addition through the 
diagonal edges $[xy]$. A loop that runs through $<xy>$ or $[xy]$ from 
$x$ to $y$ contributes
$1$ to $k_{<xy>}$ or $k_{[xy]}$ and hence $-1$ to $k_{<yx>}$ or $k_{[yx]}$.

In the worm algorithm, the ensemble is enlarged. In addition to the 
partition function as given above, there are unnormalized two-point 
functions:
\begin{equation}
 Z_{u,v} = 
\int \mbox D[\{\vec{s}\}] \exp(- {\cal H}[\{\vec{s}\}]) \vec{s}_u \cdot \vec{s}_v \;,
\end{equation}
where $\vec{s}_u \cdot \vec{s}_v = \mbox{Re} \exp(-i [\theta_u-\theta_v])$.
In the following, $u$ is the tail and $v$ the head of the worm. We note
that $Z_{u,u}= Z$, since $\vec{s}_u \cdot \vec{s}_u = 1$.
We can proceed as for the partition function above. The only difference 
is that the constraint, Eq.~(\ref{const0}), is modified for 
the sites $u$ and $v$, with $u \ne v$:
\begin{equation}
\sum_{y.nn.u}  k_{<uy>} + 
\sum_{y.3n.u}  k_{[uy]} =1
\label{const1}
\end{equation}
and
\begin{equation}
\sum_{y.nn.v}  k_{<vy>} + 
\sum_{y.3n.v}  k_{[vy]} =- 1  \;.
\label{const2}
\end{equation}
In addition to the closed loops, there is now an oriented line, the worm,
with the two ends $u$ and $v$.

The enlarged ensemble is given by 
\begin{equation}
Z_{\sum} =  \sum_{u,v} w(u,v) Z_{u,v} \;,
\end{equation}
where we have introduced the weights $w(u,v)>0$ following 
Ref. \cite{Ulli_Demo_09}.
Below we shall demonstrate that a suitable
$w(u,v)$ allows to obtain precision results 
even for large values of $q$.
Let us write $Z_{\sum}$ as a sum over configurations:
\begin{eqnarray}
Z_{\sum} = 
\sum_{u,v} w(u,v) \sum_{\{k\}'}  
\left [\prod_{<xy>} I_{k_{<xy>}}(K_1) \right]
\left [\prod_{[xy]} I_{k_{[xy]}}(K_3) \right] \;.
\end{eqnarray}
A configuration consists of the location of $u$ and $v$ 
and a choice $\{k\}'$ of 
$\{k\}$ that satisfies the constraints 
(\ref{const0},\ref{const1},\ref{const2}). 
Let us recall the two steps, 
following Ref. \cite{Ulli_Demo_09}, of the worm update:
\begin{itemize}

\item
Pick one of the neighbors of $v$ with equal probability for each of the
neighbors as proposal $v'$ for the new head site. For our model, there
are 14 possible choices. Along with that 
$k_{<v,v'>}' = k_{<v,v'>} + 1$ or 
$k_{[v,v']}' = k_{[v,v']} + 1$ is taken.

The site $v'$ along with the $k'$ is accepted with the probability 
\begin{equation}
\label{pacc}
P_{acc} =\mbox{min} [1,w(u,v') I_{k'}(K_i)/(w(u,v) I_k(K_i))] \;,
\end{equation}
where $i=1$ or $3$ for $v'$ nearest or next-to-next-to-nearest neighbor.

\item
If $u=v$, chose a new tail that is equal to the head accoring do some rule
that makes this step stable, where stable means that it conserves the correct
distribution of the $Z_{u,u}$. 
\end{itemize}
The first step fulfils detailed balance. In the second step we jump 
from $u=v$ to a new starting point of the worm. The simplest choice 
is to chose the new site with an equal probability for each site of the 
lattice. 
Below, our measurement requires that $u=(0,0,0)$. Hence we should 
insert tail and head at $u=(0,0,0)$ much more frequently than at other
sites of the lattice. This can be formally encoded by using the weight
$w(u,u)$. It is much larger for $u=(0,0,0)$ than for other sites.

The most simple quantity to determine in the worm algorithm is 
the two-point function
\begin{equation}
  \langle \vec{s}_x \cdot \vec{s}_y \rangle = \frac{Z_{x,y}}{Z_{x,x}}
= \frac{w(x,x) \langle \delta_{(x,y),(u,v)} \rangle}
       {w(x,y) \langle \delta_{(x,x),(u,v)} \rangle}   \;.
\end{equation}

Next let us discuss the algorithmic idea discussed in the appendix A
of  Ref. \cite{Cuomo23}. The 
elementary step of the worm update is equivalently reorganized. The authors
denote their version of the worm algorithm as continuous time (CT) update.

If we are at $v \ne u$, eventually we are moving to a new head $v'$.
For each of the nearest and 3N nearest neighbors $v'$ there is a probability 
\begin{equation}
p(v',v) = \frac{1}{14} \mbox{min} [1,w(v') I_{k'}(K_i)/(w(v) I_k(K_i))] \;
\end{equation}
to be the successor of $v$. In our case $1/14$ is the probability to be 
selected as proposal and then follows the acceptance probability.

The probability to get a new head of the worm in one step is
\begin{equation}
P_{move} = \sum_{v'.nn.v} p(v',v) + \sum_{v'.3n.v} p(v',v)  \le 1 \;.
\end{equation}
If no new head $v'$ is accepted, one tries again and so on. On average 
one stays at the site $v$ for the time 
\begin{equation}
t_v = \sum_{i=1}^{\infty} i P_{move} (1-P_{move})^{i-1} = 1/P_{move}  \;.
\end{equation}
Note that
\begin{equation}
 \sum_{i=1}^{\infty}  P_{move} (1-P_{move})^{i-1}  = 1
\end{equation}
meaning that we will move on at some time for sure.

In the algorithm, we go ahead straight away, taking into account $t_v$,
when computing observables. The new head is chosen among the $14$ candidates
in the following way: First we label the candidates by $\alpha=1, 2,..., 14$
and define $\tilde p_{0}=0$ and
\begin{equation}
\tilde p_{\alpha} =\sum_{\gamma=1}^{\alpha} p(v_{\gamma},v)/P_{move}
\end{equation}
Note that $\tilde p_{14} =1$.  Now $\alpha$ is selected in the following way:
We draw a random number $r$ that is uniformly distributed in $[0,1)$.
Then $\alpha$ is given by 
\begin{equation}
\tilde p_{\alpha-1} \le r < \tilde p_{\alpha}  \;.
\end{equation}

We have to evaluate $p(v_{\gamma},v)$ for all 14 neighbors. However, only
one pseudo random number is needed.
On the other hand, for the original version of the update, we need
on average $t_v$ times two random numbers along with the evaluation of 
$p(v',v)$. 

How these two choices compare, depends on the CPU times that are 
needed for the generation of a random number, the evaluation of 
$p(v',v)$ and the values of $p(v',v)$.
In our case we find a speed up by factor slightly smaller 
than two by using the version of Ref. \cite{Cuomo23}.

The question is how to deal with the insertions of the tail site.
Here in our simulations, we left the update step in its original 
form, when the worm is newly inserted, and only used the CT
version for $v \ne u$.  This way, for updating the $\{k\}$, 
we can chose a new site for the tail. 
The authors of Ref. \cite{Cuomo23} just keep the tail at $u=0$ and let
the head run forever. This is a viable choice.

\subsection{Ratios $R_q(r)=C_{q}(r)/C_{q-1}(r)$}
In principle the two-point functions $C_{q}(r)$ can be computed directly 
using the worm algorithm, by using heads and tails with the corresponding 
charges. However, the authors of \cite{Banerjee18} observed that the 
estimator of $C_{q}(r)$ becomes increasingly noisy with increasing $q$. 
The way out proposed by them is a 
divide and conquer strategy: Instead of computing $C_{q}(r)$ directly, 
ratios $R_q(r)=C_{q}(r)/C_{q-1}(r)$ are computed. To this end one simulates
\begin{equation}
Z_{\sum}  = \sum_v Z_0 \exp[-i (q-1) (\theta_0 - \theta_r)]
                  \exp[-i (\theta_0 - \theta_v)] \;.
\end{equation}
using the worm algorithm.

The simulation is started by preparing a $\{k\}$ configuration with 
charge $-q+1$ at $x=0$ and $q-1$ at $r$. In the case of $r=(L/2,0,0)$ we 
could for example set 
\begin{equation}
k_{<xy>} = q-1
\end{equation}
for all $x=(x_0,0,0)$, $y=(x_0+1,0,0)$, where $x_0 \in \{0,1,2,...,L/2-1\}$. 
For all other $<xy>$ we set $k_{<xy>} =0$. 
Obviously other more complicated paths could be taken
or the charge could be distributed over many different paths between $0$ and 
$r$. In the case of the 3N model we could also use diagonal edges to this end.
Here, for $r=(L/2,L/2,L/2)$ we used 
$<(x_0,x_0,x_0),(x_0+1,x_0,x_0)>$, 
$<(x_0+1,x_0,x_0),(x_0+1,x_0+1,x_0)>$, and 
$<(x_0+1,x_0+1,x_0),(x_0+1,x_0+1,x_0+1)>$ with
$x_0 \in \{0,1,2,...,L/2-1\}$.  

The updates of the system are done in the same way as for $q=0$. 
However, for measuring $R_{q}(r)$ we have to set the tail of the worm to 
the site $(0,0,0)$. The ratio we are looking for is obtained by
\begin{equation}
R_{q}(r)=\frac{w(0,0)}{w(0,r)} \langle \delta_{v,r} \rangle_{q-1}  \;.
\end{equation}
This means, we just have to count, how often the head of the worm visits the 
site $r$.  In the case of the continuous time version we have to add $t_r$, 
when the site $r$ is visited.

As non-trivial choice of the weight we used
\begin{equation}
w(0,x) \approx R_q^{-1}(x)=\langle \vec{s}_0 \cdot \vec{s}_x \rangle_{q-1}^{-1}  \;.
\end{equation}
The idea is that the head of the worm visits the sites of the lattice with 
equal probability for each of the sites.
We did not use an analytic expression for the two-point function, but
used preliminary numerical estimates instead. For small
values of $q$, we run the worm algorithm using $w(0,x)=1$ to this end. 
For large $q$ we iterate this procedure. As starting point we use the 
results for smaller values of $q$ that were obtained before. For the
estimate of $R_q(x)$ that is used in the production runs we performed
simulations that take order one day on a single CPU core for $L=128$.
This are about 2000000 single worm updates. We did not systematically
study the question, how many updates are optimal here. We just performed
as many updates as can be done in a reasonable time. Test runs for
smaller lattice sizes, varying the number of measurements, suggest that
we are close to optimal this way.

Sampling and storing the result, we make use of the symmetries of the lattice,
taking into account the insertions of the charges at $x=0$ and $x=r$. 
In the program we exploit
\begin{equation}
 R_q(x_0,x_1,x_2) = R_q(\pm x_0,\pm x_1, \pm x_2) \;,
\end{equation}
where the arguments are understood modulo $L$
and permutations of the components $x_0$, $x_1$ and $x_2$.

In our production runs, we first performed a certain number $n_{up}$  
of worm updates for using $w(u,x) = const$ with $u \ne (0,0,0)$ for each
measurement worm with $u=(0,0,0)$ and the non-trivial weight $w(0,x)$.
Later, in particular for large $q$, we skipped the worm 
updates with $u \ne (0,0,0)$, since this way, we still can generate 
all possible loop configurations.  For lack of time, we did not carefully
benchmark the performance of these different options.

\subsection{Gain by using non-trivial weights}
Here we jump ahead of the discussion of our simulations and already 
report the gain that is achieved  by using $w(0,x) \approx R_q^{-1}(x)$.
We define the efficiency of the update as
\begin{equation}
E_{algo}(q,L) = 1/([\mbox{CPU time needed}] \times [\mbox{stat error}]^2) \;.
\end{equation}
We have benchmarked both versions of the update, $w(0,x)=1$ and 
the non-trivial $w(0,x)=R^{-1}_q(x)$ on the same PC at
$(K_1,K_3)=(0.37069947,0.0415)$. To compare the two choices, we
compute the ratios
$r_{eff}=E_{w(0,x)=nontrivial}(q,L)/E_{w(0,x)=1}(q,L)$, which should not
depend much on the precise version of the CPU. To this end we take
the programs used in the simulation, which are reasonably well tuned.
We have computed $r_{eff}$ for a small selection of $q$ and $L$: 
For $q=2$ we get $r_{eff}=1.64$ and $2.3$ for $L=32$ and $96$, respectively.
For $q=4$ we get $r_{eff}=6.1$, $8.0$, and $10.2$ for 
$L=32$, $64$, and $96$, respectively.  In the case of $q=8$ we performed 
simulations using $w(x,0)=1$ only for small lattice sizes. For example,
we get $r_{eff}=11.7$ and $16.8$ for $L=16$ and $32$, respectively.

For given $q$, the gain by using a non-trivial weight $w(0,x)$ grows slowly
with increasing lattice size $L$. For given $L$ there is a rapid increase
with $q$. We abstain from computing the gain for larger values of $q$. 
But there can be little doubt that it further increases.

\section{Numerical results}
\label{Numerics}

In  Sec. \ref{derivatives} we discuss how we estimate the derivative of 
$R_q(L,K_1)$ with respect to $K_1$.
This derivative is needed to estimate the error in our estimates of $\Delta_q$
due to the error in $K_{1,c}$. Next in Sec. \ref{errorK3star} we discuss
leading corrections, and how the error due to residual leading corrections
to scaling at $K_{3}=0.0415$ is estimated. In Sec. \ref{Rfits} we summarize
the Ans\"atze that are used to fit the data for $R_q(L)$ obtained in 
our simulations. In Sec. \ref{standardXY} we discuss our simulations of the 
standard XY model. These are performed mainly to estimate the effect 
of leading corrections on $R_q(L)$. The results of this Sec. are needed
to estimate the error due to residual leading corrections at $K_3=0.0415$.
Then, in Sec. \ref{3Nsimulations} we discuss the simulations of the improved
3N XY model and the analysis of the $R_q(L)$ data which results in accurate
estimates of $\Delta_q$. Finally in Sec. \ref{FinalAna}, we fit our estimates 
of $\Delta_q$ to Ans\"atze based on Eq.~(\ref{CentralEQ}) to obtain
obtain estimates of the coefficients $c_{3/2}$ and $c_{1/2}$.  Next, for
$q \le 8$, we compute $D_q$ by summing up our estimates of $\Delta_q$. 
Using these estimates, we compute
$\Delta_q-c_{3/2} q^{3/2} - c_{1/2} q^{1/2}$, which is compared with the
constant part of Eq.~(\ref{CentralEQ}).

\subsection{Derivative with respect to $K_1$}
\label{derivatives}
\label{errorKc}
The derivative of $R_q(L,K_1)$ with respect to $K_1$ is needed to estimate 
the error of $\Delta_q$ due to the uncertainty of $K_{1,c}$. To this end, 
no high precision estimate is needed. 

The derivative is estimated by using the finite difference
\begin{equation}
S_{R_q}(L,K_{1,c})= 
\left . \frac{\partial R_q(L,K_1)}{\partial K_1} \right |_{K_1=0.37069947}
\approx \frac{R_q(L,K_{1,a})-R_q(L,K_{1,b})}{K_{1,a}-K_{1,b}}
\end{equation}
where $K_{1,a}>K_{1,c}$ and $K_{1,b} < K_{1,c}$ are close to the critical
coupling.
The derivative behaves as
\begin{equation}
\frac{S_{R,q}(L,K_{1,c})} {R_q(L,K_{1,c})} = a_q L^{y_t}
\end{equation}
Having determined an estimate of $a_q$, we obtain rough estimates
of $R_q(L,K_1)$ in the neighborhood of $R_q(L,K_{1,c})$ by using 
\begin{equation}
\label{K1shift}
R_q(L,K_1 + \Delta K_1)  = R_q(L,K_1) (1 + \Delta K_1 a_q L^{y_t}) \;.
\end{equation}
Throughout, for all values of $q$, 
we simulated at $K_1=0.3700$ and $0.3714$ for the linear lattice size $L=16$.
In the case of $q=4$, as check, we varied the lattice size and the values
of $K_1$. We get consistent results.

The error induced on $\Delta_q$ is estimated by redoing fits using 
the data for $R_q$ multiplied by $1 + \Delta K_1 a_q L^{y_t}$, where
$\Delta K_1$ is the error of $K_{1,c}$ now. 

\subsection{Residual leading corrections to scaling}
\label{errorK3star}
The value of $K_3^*$ is only approximately known. Therefore there
might be a small amplitude of the leading correction at $K_3=0.0415$. 

In ref. \cite{myXY2025} we have determined $K_3^*=  0.04149(14)$ by analyzing 
dimensionless quantities. Let us briefly recall the logic behind 
using improved models in a simplified setting:  Let us take the Binder
cumulant and its slope as example. We assume for simplicity that the
critical temperature is exactly known. Then in general
\begin{equation}
 U_4(L,K_{1,c},K_3) = U_4^* + b(K_3) L^{-\omega} 
+ c [b(K_3) L^{-\omega}]^2 + ...
         + d(K_3) L^{-\epsilon} + ... \;.
\end{equation}
To further simplify the argument, we focus on the leading correction. In the
fit we are using the Ansatz
\begin{equation}
\label{U4_corr}
U_4(L,K_{1,c},K_3) = U_4^* + b(K_3) L^{-\omega}  \;.
\end{equation}
Furthermore, we assume that we know $\omega$ from other sources.
Hence, for a given value of $K_3$, the parameters of the fit are 
$U_4^*$ and $b(K_3)$. Say in our study we
are aiming at the RG exponent $y_t$. In finite size scaling, 
we extract it from the slope of dimensionless quantities. Here we take the 
Binder cumulant as example
\begin{equation}
\label{slope_corr}
 S_{U_4}(L,K_{1,c},K_3) =
\left . \frac{\partial U_4(L,K_{1},K_3)}{ \partial K_1} \right |_{K_1=K_{1,c}}
= a(K_3) L^{y_t} (1 + p b(K_3) L^{-\omega} + ...  ) \;.
\end{equation}
Say we simulate only the XY model, $K_3 =0$. Then we 
have three free parameters in the fit: $a(0)$, $y_t$, and $p b(0)$.
One has to keep in mind that the more free parameters there are in the 
Ansatz, the larger is the statistical error of the parameters in the fit. 
Hence, by fitting $U_4$ we get $b(K_3)$ more accurately than 
$p b(K_3)$ by fitting $S_{U_4}$.  

Simulating at $K_3 \approx K_3^*$, we know that $p b(K_3^*)$ is small. However,
simply ignoring the correction term in the Ansatz results in a systematic 
error in our estimate of $y_t$, which is likely small. However, we would
like to know how small. Here information from the simulation of a model
with a reasonably large leading correction amplitude, say the XY model, helps. 
Fitting the estimates of $U_4$ and its slope for the XY model, we get 
estimates of $b(K_3)$ and $p b(K_3)$ and hence $p=[p b(K_3)]/b(K_3)$.  

Analyzing the Binder cumulant $U_4$ for the XY and the 3N XY model
at $K_3=0.0415$ and $0.046$, using the data of Refs. \cite{myClock,myXY2025} 
we  translate the error in $K_3^*=0.04149(14)$ to
the bound  $|b(0.0415)/b(0)| \lessapprox  250$.
Hence the amplitude of the leading correction in $S_{U_4}$ at 
$K_3=0.0415$ at criticality is bounded by
\begin{equation}
|p b(0.0415)| \lessapprox |p b(0)|/250 \; .
\end{equation}

Now we estimate the possible error of $y_t$ due to leading corrections 
as follows. First we fit our data for $K_3=0.0415$  with the Ans\"atze
\begin{equation}
S(L) = a L^{y_t}
\end{equation}
and for example
\begin{equation}
S(L) = a L^{y_t}  (1 + c L^{-\epsilon}) \;,
\end{equation}
where $\epsilon \approx 2$  corresponding to subleading corrections.
Here we take our data for $S(L)$ obtained in the simulations.
Next we repeat these fits, replacing the data for $S(L)$ by
the data multiplied by $1+(|p b(0)|/250) L^{-\omega}$. 
The difference between these two sets of fits provides us with an estimate
of the error due to residual leading corrections at $K_3=0.0415$.
Below, analyzing $R_q(L)$, we proceed in an analogous way.

\subsection{Fit Ans\"atze}
\label{Rfits}
In the case of the 3N XY model, we performed fits by using the Ans\"atze 
\begin{equation}
\label{fit0}
 R_q(L) = a(q) L^{-2 \Delta_q} \;,
\end{equation}
\begin{equation}
\label{fit2}
 R_q(L) = a(q) L^{-2 \Delta_q}  [1+ c(q) L^{-2}]  \;
\end{equation}
and
\begin{equation}
\label{fit4}
 R_q(L) = a(q) L^{-2 \Delta_q}  [1+ c(q) L^{-2} +d(q) L^{-4}]  \;.
\end{equation}
The correction $\propto L^{-2}$ is due to second derivatives of the 
field with respect to the location. See for example Appendix C of 
Ref. \cite{Meneses19}. There should be corrections due to the violation
of the rotational invariance by the lattice. The corresponding 
correction exponent is $\omega_{NR}=2.02548(41)$ \cite{O2corrections,private}. 
In the 3N XY model at $K_3=0.0415$ the amplitude of this 
corrections is suppressed by approximately a factor of 9 compared 
with the standard XY model \cite{myXY2025}. In the fits we expect that 
this correction is effectively taken into account by the term
$\propto L^{-2}$.
The correction $\propto L^{-4}$ is taken
ad hoc. It is mainly justified by the fact that it nicely fits the data.
Comparing the results obtained by using the different Ans\"atze provides us 
with an estimate
of the systematic error caused by higher order corrections to scaling.
It turns out that the amplitudes of corrections to scaling increase
with increasing $q$. Therefore the Ansatz~(\ref{fit4}) is useful in 
particular for larger values of $q$.  

In the case of the standard XY model we used as above Eq.~(\ref{fit0}) and
in addition
\begin{equation}
\label{fit0om}
 R_q(L) = a(q) L^{-2 \Delta_q} [1+ b(q) L^{-\omega}]  \;
\end{equation}
and
\begin{equation}
\label{fit2om}
 R_q(L) = a(q) L^{-2 \Delta_q}  [1+ b(q) L^{-\omega} + c(q) L^{-2}]  \;
\end{equation}
to take leading corrections to scaling into account.

\subsection{The standard XY model}
\label{standardXY}
Since the leading correction is not exactly eliminated in the 
3N XY model, we need simulations of the standard XY to estimate 
the residual amplitude of the leading correction in the 3N XY model.
We simulated the standard XY model at $K=K_1=0.45416475$. 
Note that
$K_{1,c} =1/2.2018441(5) \approx 0.45416476(10)$
given in Ref. \cite{Deng19}  and $K_{1,c} =0.45416474(10)[7]$ \cite{myClock}.
Here we simulated for $q=2$, $3$, $4$, $6$, $8$, $16$ and $32$. 
The choice of linear lattice sizes and the statistics is similar 
as above.  

\begin{figure}
\begin{center}
\includegraphics[width=14.5cm]{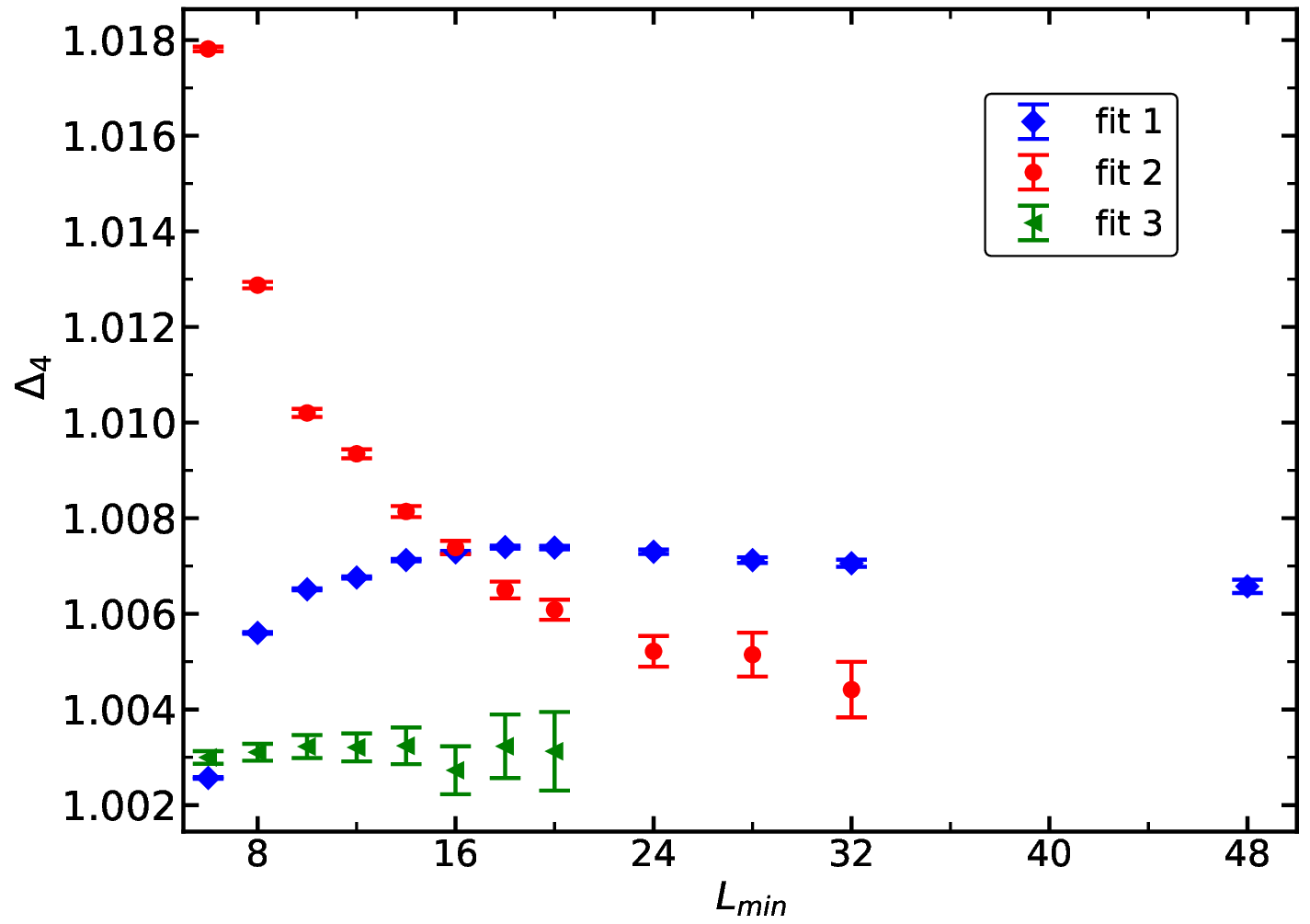} 
\caption{\label{DeltaXYQ4}
Data for $R_4(L)$ for the standard XY model at $K=0.45416475$ are analyzed.
Estimates of $\Delta_4$ are plotted versus the minimal linear lattice
$L_{min}$ that is taken into account in the fit. The estimates are 
obtained by using the Ans\"atze (\ref{fit0},\ref{fit0om},\ref{fit2om}).
In the caption we denote the fits by fit 1, fit 2, and fit 3, respectively.
Details are discussed in the text.
}
\end{center}
\end{figure}

As an example let us discuss the analysis of our data for $q=4$ in 
detail. We simulated the linear lattice sizes $L=6$, $8$, ..., $20$, $24$,
..., $32$, $48$, $64$, and $96$. 
We had started with simulations with $w(0,x)=1$
and had reached a statistics that we regarded as final. Later, having 
realized the gain that could be achieved by the proper choice of 
$w(0,x)$, we have redone the simulations, now achieving a better accuracy.
Using $w(0,x) \approx R_q^{-1}(x)$, we performed 
for example $10^{10}$ measurements for $L=16$. This number decreases to 
$7.3 \times 10^8$ for $L=96$. The simulations took about the equivalent of 
1.1 years 
of CPU time on a single core of an Intel(R) Xeon(R) CPU E3-1225 v3 
running at 3.20 GHz for $w(0,x) \approx R_q^{-1}(x)$ in total.
We checked that the results obtained in these two sets of simulations
are consistent. We merged the data before
fitting.  In FIG. \ref{DeltaXYQ4} we plot the estimates of $\Delta_q$ 
obtained by using the Ans\"atze~(\ref{fit0},\ref{fit0om},\ref{fit2om}). 
Note that we get fits with  $\chi^2/$DOF $\approx 1$
for the Ansatz (\ref{fit2om}) starting from $L_{min} =6$. In the case of
the Ansatz~(\ref{fit0om}), we get $\chi^2/$DOF $=1.73$ for $L_{min} =20$ but
it does not further drop. For the pure power law, Eq.~(\ref{fit0}), we 
get $\chi^2/$DOF $=9.87$ for $L_{min} =18$ but $\chi^2/$DOF does not 
further decrease towards clearly smaller values, when increasing 
$L_{min}$. In our fits all linear lattice sizes $L \ge L_{min}$ are taken 
into account.
Fitting with the Ansatz~(\ref{fit2om}) gives a stable result starting 
from the smallest lattice sizes considered here. Here we are not quoting 
a final estimate. To this end, we should take into account that there 
are corrections proportional to $L^{-2 \omega}$, which are not included
into the Ansatz. Also the error due 
to the uncertainty of the correction exponent $\omega$ should be 
taken into account in the estimate of the systematic error. But let 
us keep in mind that for example for $L_{min}=12$, 
using the Ansatz~(\ref{fit2om}), we get 
$\Delta_4=1.0032(3)$ just to see how it compares with the estimate
obtained from the improved model.
In the following, we need an estimate of $b(4)$,
Eqs.~(\ref{fit0om},\ref{fit2om}), to estimate the error due to residual
leading corrections in the case of the 3N XY model. 
Based on the Ansatz~(\ref{fit2om}) we get $b(4) = 0.27(7)$.
For the other values of $q$ we proceeded in a similar fashion.  We find 
$b=0.00(6)$, $0.18(10)$, $0.55(9)$, $0.74(6)$, $1.56(6)$, $3.3(1)$, 
for $q=2$, $3$, $6$, $8$, $16$, and $32$, respectively. We note 
that the amplitude of corrections vanishes within error bars for $q=2$.
Roughly it increases linearly with increasing $q$. 
Below in Sec. \ref{3Nsimulations} we shall use estimates based on a linear fit
$b(q) =-0.165+0.1087 q$
for the values of $q$ that we did not simulate for the standard XY model.

\subsection{3N XY model}
\label{3Nsimulations}
As discussed in Sec. \ref{themodel}, we simulated the 3N XY model at
$(K_1,K_3)=(0.37069947,0.0415)$.
For the 3N XY model, we performed simulations for  $q=2$, $3$, ..., $8$, 
$16$, $24$, $32$, $48$ and $64$. For $q = 2$ and $3$, 
we simulated the linear lattice sizes $L=6$, $8$, $10$, $12$, $14$, $16$, $18$,
$20$, $24$, $28$, $32$, $48$, $64$, $96$ and $128$. 
For $q=4$, $5$, ..., $8$ we omit $L=128$. 
For $q=16$, $24$ and $32$ we start from $L=12$ and simulate up to 
$L=128$,
while for $q=48$ and $64$ we start from $L=16$ and simulate up to $L=128$.
The lattice sizes in between are chosen as above for $q = 2$ and $3$.
For small $q$, we have generated
data using $w(0,x)=1$ before switching to non-trivial $w(0,x)$.  
We checked that the results obtained in these two sets of simulations
are consistent. We merged the data before fitting.

For $q=2$ and the non-trivial $w(0,x)$, we performed about $4 \times 10^{10}$
measurements for $L=16$. This number decreases to $1.7 \times 10^9$ for
$L=128$. For example for $L=128$ we get $R_2(128) = 0.00048747(10)$.
For larger $q$ the statistics is lower.
The simulations for $q=2$ and $3$ took the equivalent of about 
3.3 and 2.5 years of CPU time on a single core of an
Intel(R) Xeon(R) CPU E3-1225 v3 running at 3.20 GHz
for $w(0,x) \approx R_q^{-1}(x)$ in total, respectively.
For $q \ge 4$, we used slightly more than one year of CPU time for each value 
of $q$.

\begin{figure}
\begin{center}
\includegraphics[width=11.0cm]{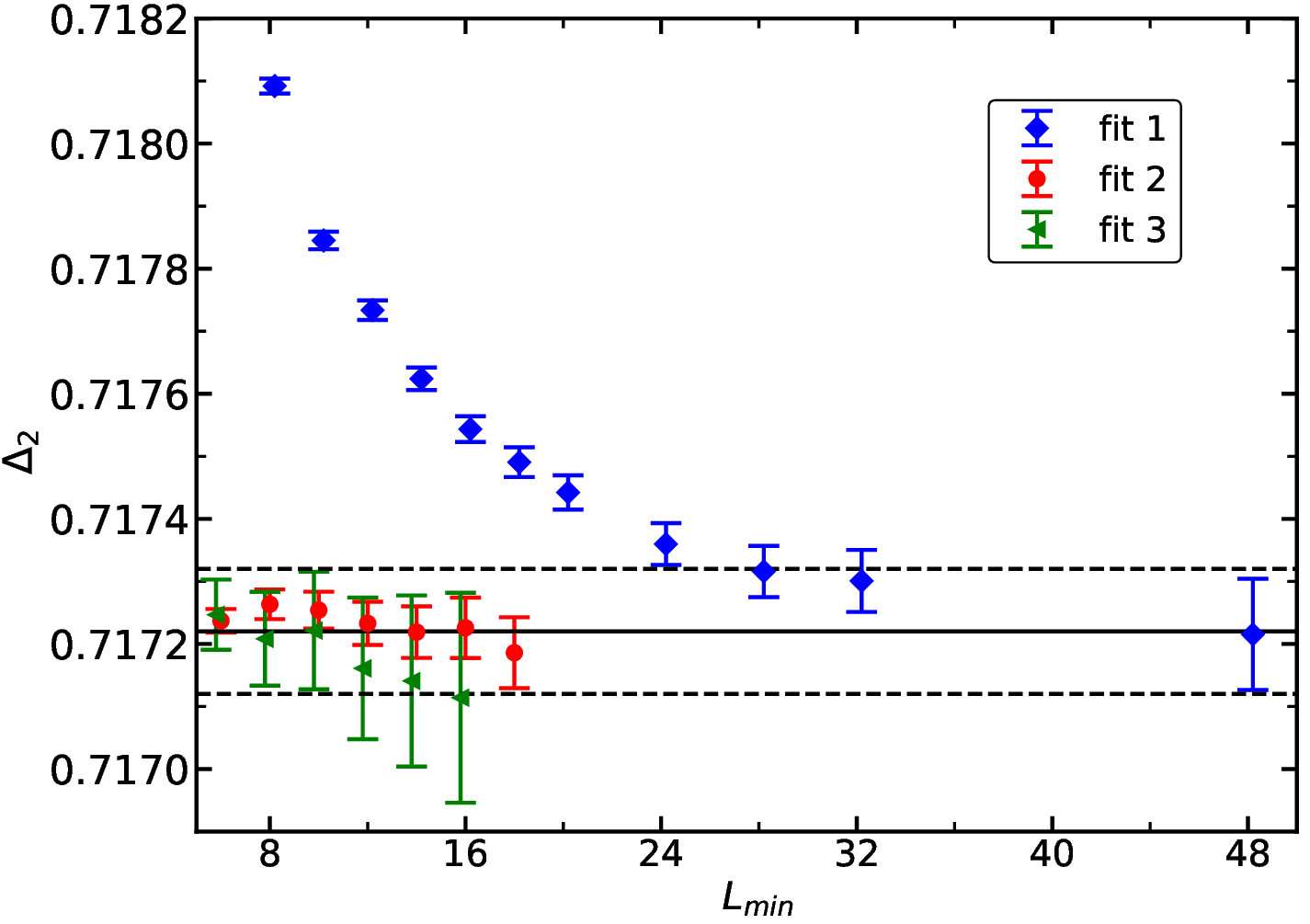}
\caption{\label{DeltaQ2}
Data for the 3N XY model at $K_1=0.37069947$ and $K_3=0.0415$ are analyzed.
Estimates of $\Delta_2$ obtained by fitting  $R_2(L)$
using the Ans\"atze (\ref{fit0},\ref{fit2},\ref{fit4}) are plotted versus the
minimal linear lattice size $L_{min}$ that is taken into account.
In the caption we denote the fits by fit 1, fit 2, and fit 3, respectively.
The solid line indicates our final estimate, while the dashed lines give the
error.
The values of $L_{min}$ are slightly shifted to avoid overlap of the symbols.
Details are discussed in the text.
}
\vskip1.0cm
\includegraphics[width=11.0cm]{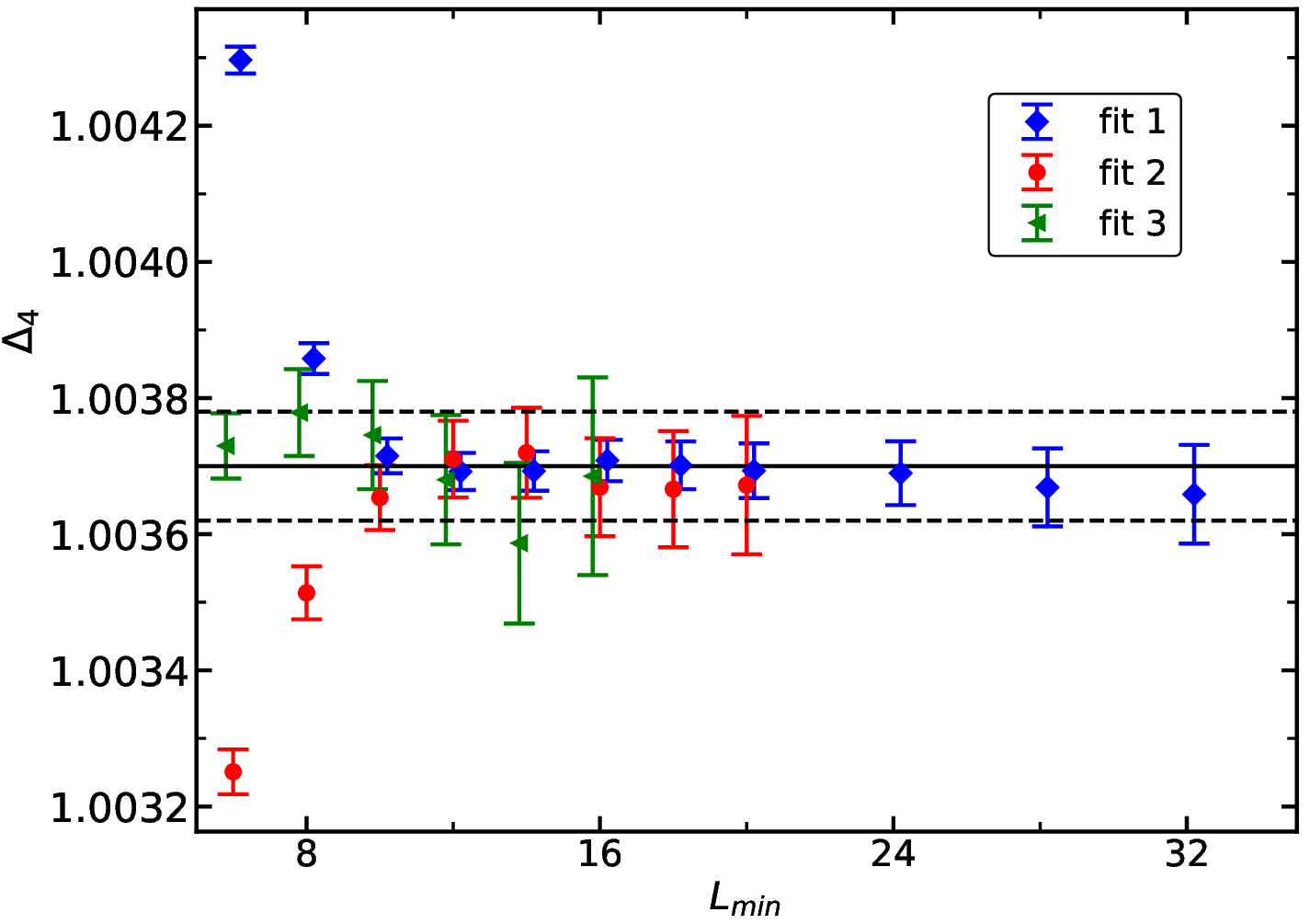}
\caption{\label{DeltaQ4}
Same as FIG. \ref{DeltaQ2} but for $q=4$ instead of $q=2$.
}
\end{center}
\end{figure}

\begin{figure}
\begin{center}
\includegraphics[width=11.0cm]{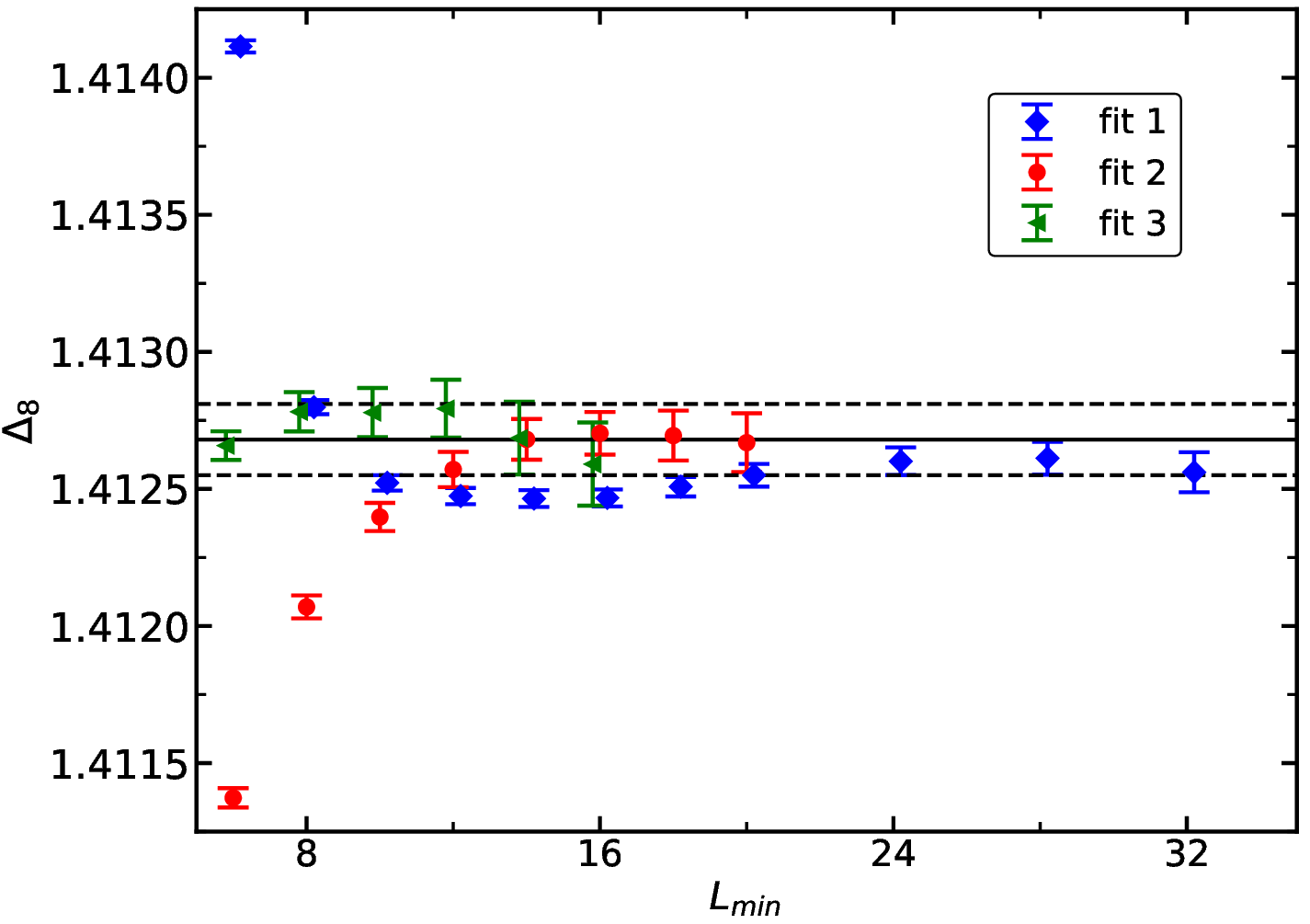}
\caption{\label{DeltaQ8}
Same as FIG. \ref{DeltaQ2} but for $q=8$ instead of $q=2$.
}
\vskip1.0cm
\includegraphics[width=11.0cm]{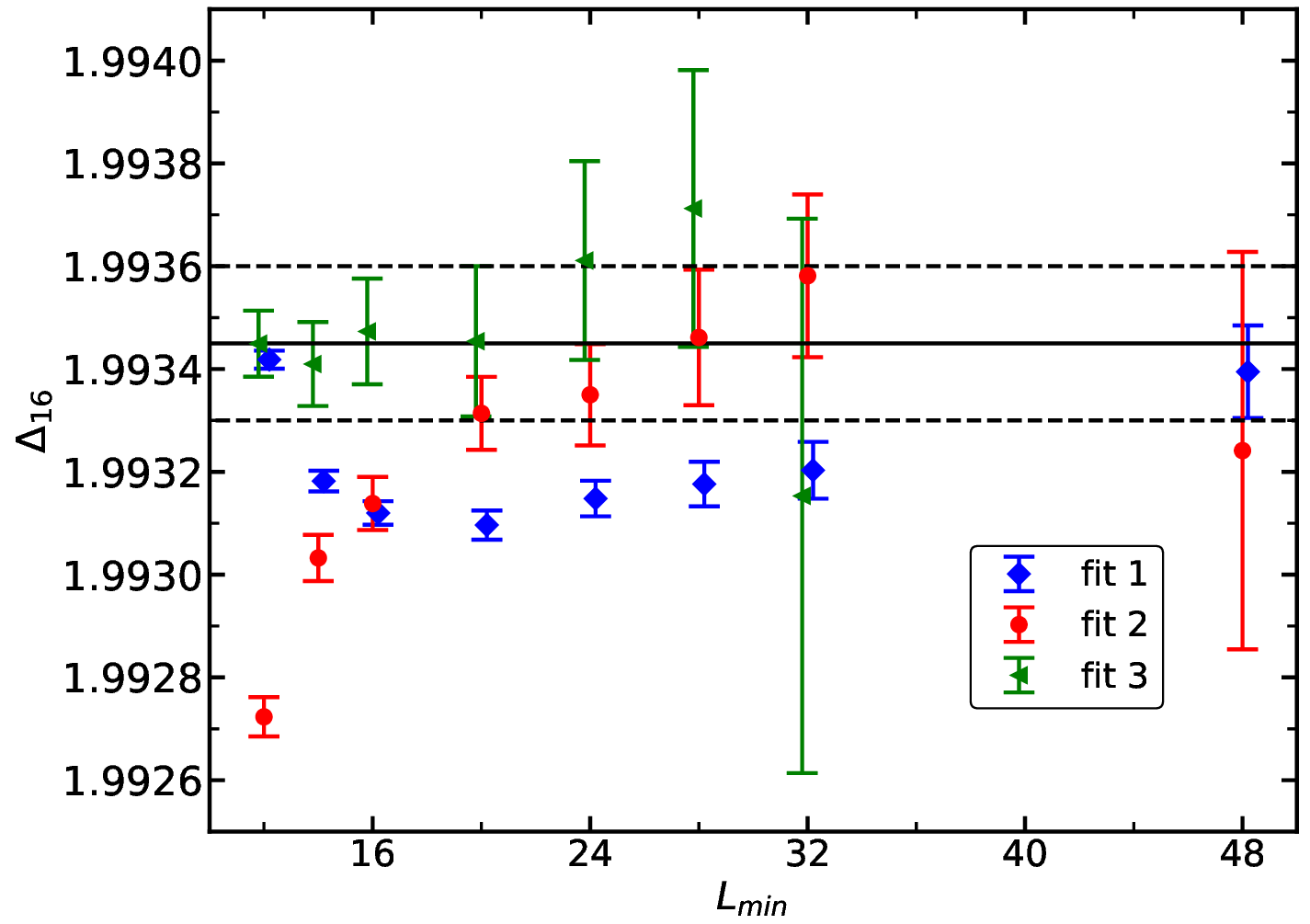}
\caption{\label{DeltaQ16}
Same as FIG. \ref{DeltaQ2} but for $q=16$ instead of $q=2$.
}
\end{center}
\end{figure}

\begin{figure}
\begin{center}
\includegraphics[width=11.0cm]{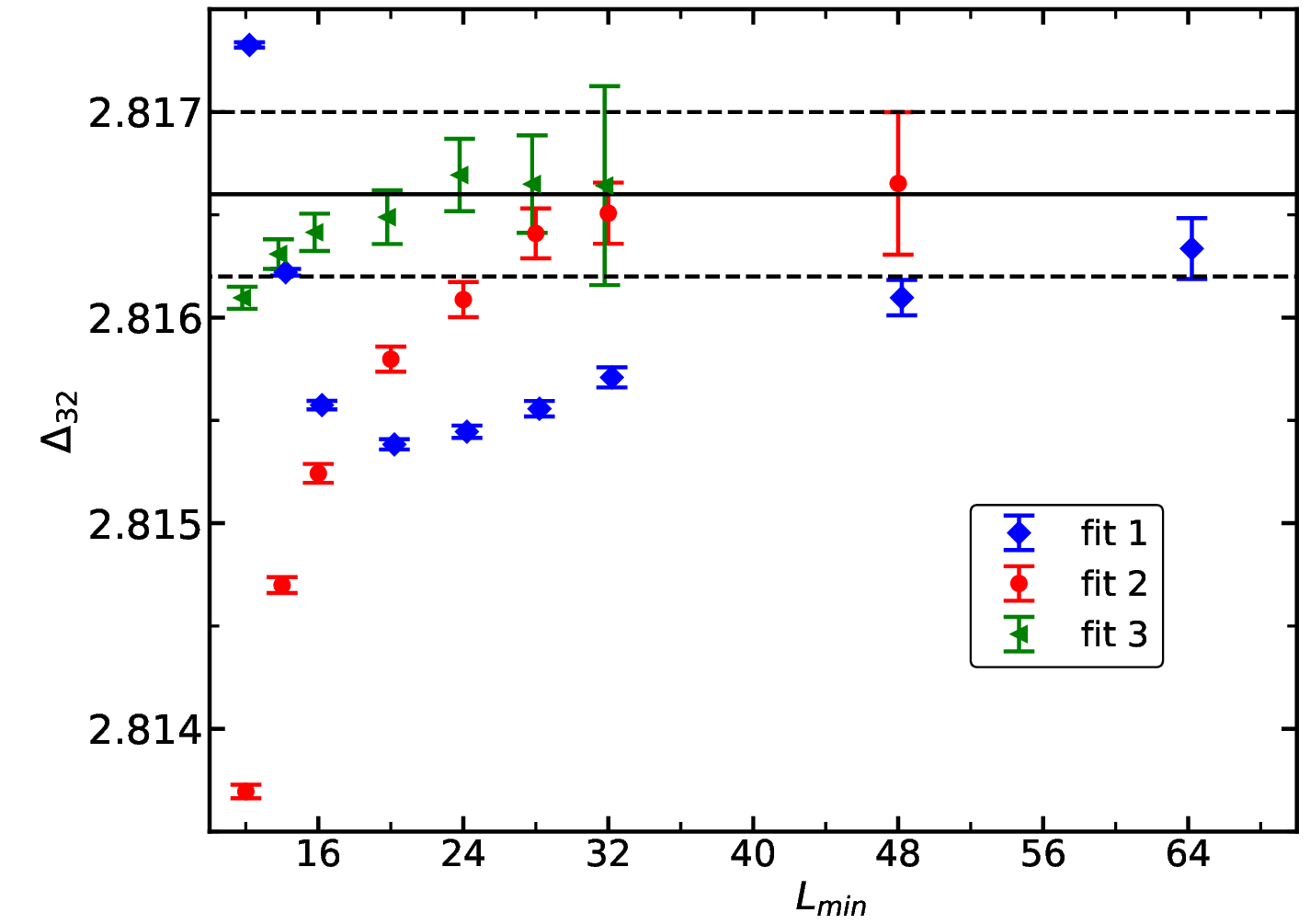}
\caption{\label{DeltaQ32}
Same as FIG. \ref{DeltaQ2} but for $q=32$ instead of $q=2$.
}
\vskip1.0cm
\includegraphics[width=11.0cm]{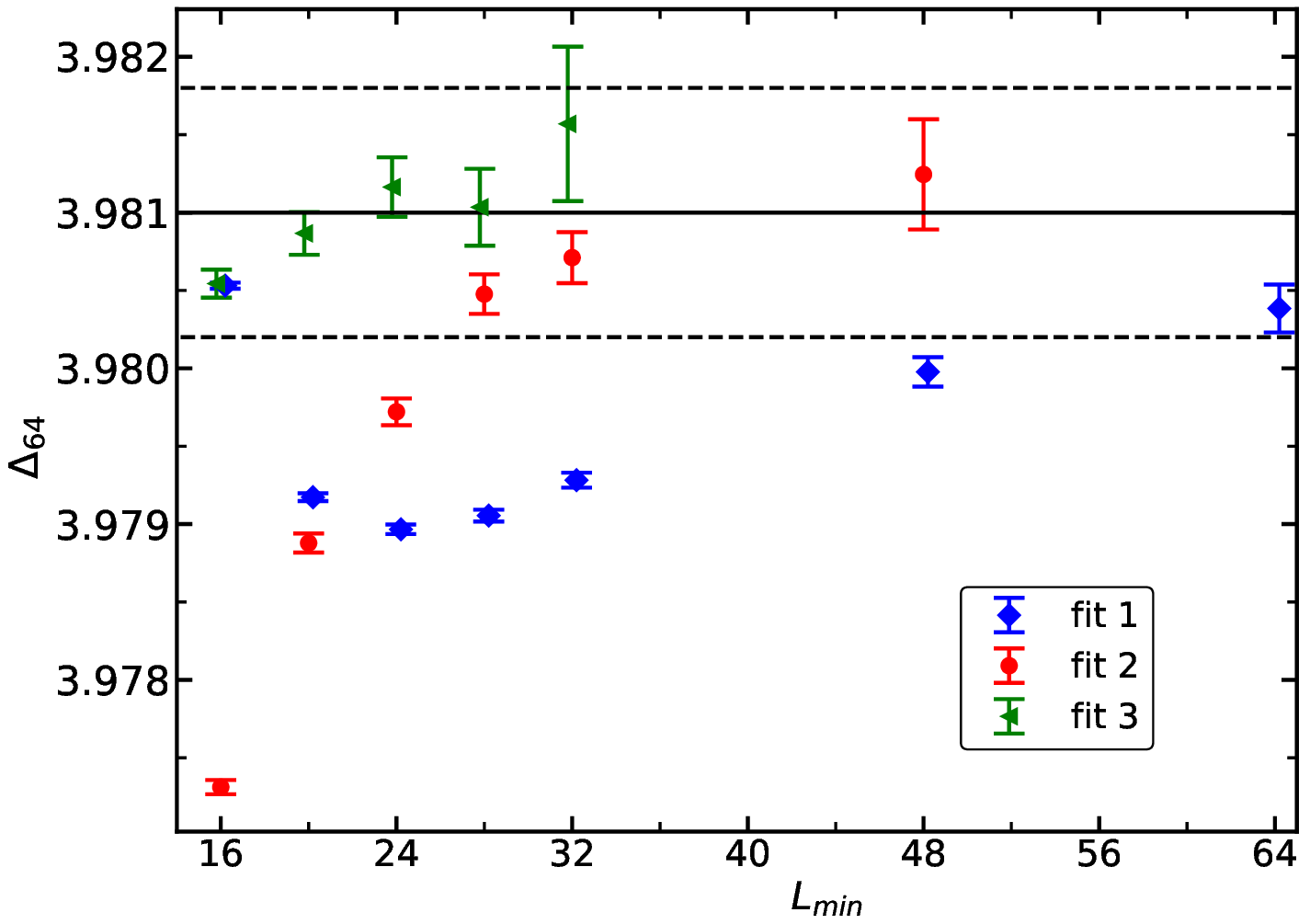}
\caption{\label{DeltaQ64}
Same as FIG. \ref{DeltaQ2} but for $q=64$ instead of $q=2$.
}
\end{center}
\end{figure}

In FIGs. \ref{DeltaQ2}, \ref{DeltaQ4}, \ref{DeltaQ8}, \ref{DeltaQ16}, 
\ref{DeltaQ32}, and \ref{DeltaQ64} we plot our results for $\Delta_q$ obtained
by fitting $R_q(L)$ with the Ans\"atze~(\ref{fit0},\ref{fit2},\ref{fit4}) for 
$q=2$, $4$, $8$, $16$, $32$, and $64$, respectively, versus $L_{min}$.  
Note that in these
fits all linear lattice sizes with $L \ge L_{min}$ are taken into account.
Typically in such fits $\chi^2/$DOF has a large value for small $L_{min}$
and decreases rapidly with increasing $L_{min}$ reaching 
$\chi^2/$DOF $\approx 1$. In table \ref{Goodness} we give the $L_{min}$, 
where this point is 
reached depending on $q$ and the Ansatz that is used. Let us denote this
minimal lattice size by $L_{min,min}$. Let us denote fits with 
$L_{min} \ge L_{min,min}$ as acceptable fits. Note that this says little
on systematic errors on the parameters of the fit due to corrections that
are not taken into account in the Ansatz. In order to get some handle 
on systematic errors, we require that our final estimate is consistent 
with fits using different Ans\"atze. The final estimate is chosen such
that it covers for each of the Ans\"atze~(\ref{fit0},\ref{fit2},\ref{fit4})
at least the estimate obtained by one acceptable fit.
By $q \pm \Delta q$ covers
$p \pm \Delta p$ we mean that $q + \Delta q \ge p + \Delta p$ and
$q - \Delta q \le p - \Delta p$ hold, where $\Delta p$ is the error of the
parameter $p$. In the FIGs. our final result is shown as solid line and 
the error is indicated by dashed lines.  Our final estimates are summarized in 
table \ref{FinalDelta}.

We note that $L_{min,min}$ increases with increasing $q$ for all 
Ans\"atze that we use. The only exception is the Ansatz~(\ref{fit0}), 
where we have a minimum at $q=4$. Going to large values of $q$, 
the amplitudes of corrections increase rapidly. This can also be seen
directly by looking at the results obtained by using the 
Ans\"atze~(\ref{fit2},\ref{fit4}).

\begin{table}
\caption{\sl \label{Goodness}
We give $L_{min,min}$ as a function
of $q$ and the Ansatz. For the definition of $L_{min,min}$  see the text.
The values of $q$ correspond to the FIGs. 
\ref{DeltaQ2}, \ref{DeltaQ4}, \ref{DeltaQ8}, \ref{DeltaQ16},
\ref{DeltaQ32}, and \ref{DeltaQ64}.
Note that for $q=2$, $4$, and $16$ the lattice sizes $L=6$, $6$ and $12$
are the smallest that we have simulated.  For a discussion see the text.
}
\begin{center}
\begin{tabular}{ccccc}
\hline
$q$ & \textbackslash{} Ansatz  & ~(\ref{fit0}) & ~(\ref{fit2}) & 
~(\ref{fit4})  \\
\hline
 2  &  &    24     &    8    &    6  \\
 4  &  &    12     &   10    &    6   \\
 8  &  &    24     &   14    &    8   \\
16  &  &    48     &   20    &   12   \\
32  &  &    64     &   28    &   16   \\
64  &  &    -      &   48    &   24   \\
\hline
\end{tabular}
\end{center}
\end{table}

Estimating the systematic errors that are due to the uncertainty in 
$K_{1,c}$ and $K_3^*$, we follow the discussions of Secs. \ref{errorKc} and
\ref{errorK3star}. The corresponding errors are given in table \ref{FinalDelta}.
We note that the error due to the uncertainty in $K_{1,c}$ is roughly the same
for all values of $q$, while the error due to the uncertainty in  $K_3^*$
increases with increasing $q$.

\begin{table}
\caption{\sl \label{FinalDelta}
Our final estimate of $\Delta_q$ for all values of $q$ we simulated at. 
In $()$ we give the error assuming that we are exactly at the critical coupling
and that the amplitude of leading corrections is exactly vanishing.
$\delta K_{1,c}$ is the possible error due to the uncertainty of 
our estimate of $K_{1,c}$  and $\delta K_{3}^*$ the possible error due to the
uncertainty of our estimate of $K_{3}^*$. For a discussion see the text.
}
\begin{center}
\begin{tabular}{clcl}
\hline
$q$ &\mc{1}{c}{$\Delta_q$} & $\delta K_{1,c}$ & \mc{1}{c}{$\delta K_{3}^*$} \\
\hline
2   & 0.71722(10) & 0.000008 & 0.000005 \\
3   & 0.87203(8)  & 0.000008 & 0.00002 \\
4   & 1.00370(8)  & 0.000006 & 0.00003 \\
5   & 1.11992(9)  & 0.000006 & 0.00003 \\
6   & 1.22519(10) & 0.000006 & 0.00005 \\
7   & 1.32220(10) & 0.000006 & 0.00005 \\
8   & 1.41268(13) & 0.000006 & 0.00006 \\
16  & 1.99345(15) & 0.000009 & 0.00008 \\
24  & 2.4399(3)   & 0.000009 & 0.00012 \\
32  & 2.8166(4)   & 0.000009 & 0.00015 \\
48  & 3.4482(5)   & 0.000008 & 0.00024 \\
64  & 3.9810(8)   & 0.000008 & 0.00032 \\
\hline
\end{tabular}
\end{center}
\end{table}

\subsection{Fitting by the large $q$-expansion}
\label{FinalAna}
Starting from Eq.~(\ref{CentralEQ}) we fitted our data for 
$\Delta_q$ by using the Ans\"atze
\begin{equation}
\label{DeltaFit1}
 \Delta_q = c_{3/2} [q^{3/2}-(q-1)^{3/2}] + 
               c_{1/2} [q^{1/2}-(q-1)^{1/2}] 
\end{equation}
and
\begin{equation} 
\label{DeltaFit2}
 \Delta_q = c_{3/2} [q^{3/2}-(q-1)^{3/2}] + 
               c_{1/2} [q^{1/2}-(q-1)^{1/2}] +
               c_{-1/2} [q^{-1/2}-(q-1)^{-1/2}]  \;.
\end{equation}
Since the error is partially 
systematic and furthermore the systematic error might point in the 
same direction for all values of $q$, $\chi^2/$DOF has to be taken with 
caution. 
The same holds for the errors of the parameters of the fit.
In the fit we only used the error, assuming that we are exactly at the critical
temperature and that leading corrections to scaling vanish exactly.

In table 
\ref{coefficients2}  we have summarized results of fits by using the 
Ansatz~(\ref{DeltaFit2}), where all $\Delta_q$ with $q \ge q_{min}$ 
are taken into 
account. We see that $\chi^2/$DOF makes a jump going from $q_{min}=3$ 
to $4$. $\chi^2/$DOF for $q_{min}=4$ is very small. Likely this is due to the 
fact that the systematic error in our estimates of $\Delta_q$ points in the 
same direction for different values of $q$. We also note that the values
of $c_{3/2}$, $c_{1/2}$  change only little going 
from $q_{min}=4$ to $5$. 
It is very remarkable that the 
estimates of $c_{-1/2}$ are very small and that they are decreasing 
with increasing $q_{min}$. For $q_{min}=5$ the estimate of $c_{-1/2}$ 
is compatible with zero.  It seems plausible that $c_{-1/2}$ is very 
tiny or even exactly zero and deviations from Eq.~(\ref{DeltaFit1}) 
are due to higher powers or even decay exponentially in $q$.

Hence we performed fits by using the Ansatz~(\ref{DeltaFit1}). 
The results are summarized on the lower half of table \ref{coefficients2}. 
Here  $\chi^2/$DOF is still large for $q_{min}=3$. However, very small already
for $q_{min}=5$. We take our final estimates of $c_{3/2}$ and $c_{1/2}$
from this fit. Going to larger $q_{min}$ the statistical error increases, 
likely without gaining much on systematic errors.

\begin{table}
\caption{\sl \label{coefficients2}
We fit our data for $\Delta_q$ by using the Ans\"atze
(\ref{DeltaFit1},\ref{DeltaFit2}). $c_{3/2}$, $c_{1/2}$, and $c_{-1/2}$
are the coefficients of Eq.~(\ref{CentralEQ}).
}
\begin{center}
\begin{tabular}{ccclll}
\hline 
 $q_{min}$ & Ansatz  &  $ c_{3/2} $ &  \mc{1}{c}{$c_{1/2}$} & $c_{-1/2}$  & $\chi^2/$DOF \\
\hline 
   2  &  \ref{DeltaFit2}  &  0.331553(22) & 0.2815(5) & 0.0191(9) & 0.62 \\
   3  &  \ref{DeltaFit2}  &  0.331566(27) & 0.2809(9) & 0.0173(22) & 0.60 \\
   4  &  \ref{DeltaFit2}  &  0.331598(31) & 0.2787(14)& 0.0095(45) & 0.12 \\
   5  &  \ref{DeltaFit2}  &  0.331602(37) & 0.2784(22)& 0.0081(89) & 0.13 \\
\hline 
   3  &  \ref{DeltaFit1}  &  0.331725(18) & 0.27389(27) & \mc{1}{c}{-}& 7.75 \\
   4  & \ref{DeltaFit1} &  0.331648(21) & 0.27587(37)& \mc{1}{c}{-}& 0.66 \\
   5  &  \ref{DeltaFit1} &  0.331628(23) & 0.27648(49)&\mc{1}{c}{-}&  0.23 \\
   6  &  \ref{DeltaFit1} &  0.331623(25) & 0.27670(62)& \mc{1}{c}{-}&0.21 \\
\hline 
\end{tabular}
\end{center} 
\end{table}

\begin{eqnarray}
\label{finalC32}
c_{3/2} &=& 0.33163(9) \\  
\label{finalC12}
c_{1/2} &=& 0.2765(8)  
\end{eqnarray}
The error is due to different sources. First we take the error 
given in table \ref{coefficients2}. Then, assuming that the error
is purely systematic and points in the same direction for all values of $q$, 
we have redone the fit using $\Delta_q+\epsilon$, where $\epsilon$ is the 
error that is quoted in table \ref{FinalDelta} in $()$.  The error is then 
the difference between fitting  $\Delta_q+\epsilon$ and $\Delta_q$. 
Analogously, we proceeded for $\Delta_q+\delta K_{1,c}$ and 
$\Delta_q+\delta K_{3}^*$. For $c_{3/2}$  these errors are 
$0.000023$, $0.000042$, $0.000002$, and $0.000019$, respectively. 
In the case of 
$c_{1/2}$ we get $0.00049$, $0.00024$, $0.000008$, and $0.00011$, respective. 
To get the errors in Eqs.~(\ref{finalC32},\ref{finalC12}), 
we just added up these four numbers. 
We regard these error estimates as cautious.

Next, we add up the estimates of $\Delta_q$, where we start from 
$D_1 = 0.51908(1)$ \cite{myXY2025}. We just sum up the errors, 
assuming that they are systematic.  The results are given in table 
\ref{Dqsummed}. To compare with the constant $-0.0937256$ in 
Eq.~(\ref{CentralEQ}) we compute $D_q -c_{3/2} q^{3/2} - c_{1/2} q^{1/2}$.
The difference seems to converge rapidly with increasing $q$. For
$q=5$, $6$, and $7$ it takes the value $-0.0940$ which is very close
to the constant that is predicted analytically. The small difference can 
be explained by the error on $D_q$ and the coefficients $c_{3/2}$ and
$c_{1/2}$. It is remarkable that the difference even for $q=1$ is just 
about $0.005$. 

\begin{table}
\caption{\sl \label{Dqsummed}
Estimates of $D_q$ obtained by summing up $\Delta_q$. The starting point is
$D_1 = 0.51908(1)$ \cite{myXY2025}.  To compare with the constant in 
Eq.~(\ref{CentralEQ}) we compute $D_q -c_{3/2} q^{3/2} - c_{1/2} q^{1/2} $. 
}
\begin{center}
\begin{tabular}{ccclll}
\hline
$q$ & $D_q$   &  $D_q -c_{3/2} q^{3/2} - c_{1/2} q^{1/2} $ \\
\hline
 1  & 0.51908(1)\phantom{0} & --0.0890 \\
 2  & 1.23630(12)& --0.0927 \\
 3  & 2.10833(23)& --0.0937 \\
 4  & 3.11203(34)& --0.0939 \\
 5  & 4.23195(47)& --0.0940 \\
 6  & 5.45714(62)& --0.0940 \\
 7  & 6.77934(78)& --0.0940 \\
 8  & 8.19202(97)& --0.0939 \\
\hline
\end{tabular}
\end{center}
\end{table}

\section{Summary and outlook}
\label{summary}
We have determined the critical exponents $Y_q=3-D_q$ of a $\mathbb{Z}_q$ 
invariant perturbation at the $O(2)$ invariant fixed point in 
three dimensions by using Monte Carlo simulations of an improved
model. In particular, we studied the XY model with next-to-next-to-nearest
neighbor couplings in addition to nearest neighbor couplings at the 
point, where leading corrections to scaling approximately vanish \cite{myXY2025}.
We simulate the model by using the worm algorithm \cite{Prok98,Prok01}.
The worm algorithm has been used to simulate the XY model in several
previous works. See for example \cite{Banerjee10,Deng19}. It is
straight forward to apply the algorithm to the model studied
here.

We build on previous work \cite{Banerjee18,Cuomo23} on this problem.
The dimensions $D_q$ are obtained in an iterative approach.
Considerable improvement is achieved by using site weights
\cite{Ulli_Demo_09} in
the worm simulation. The gain obtained by using this modification is increasing
with increasing $q$. In particular, this allows us to reach $q=64$
at high precision.
The authors of \cite{Banerjee18,Cuomo23} simulated the standard XY model. 
By simulating an improved
model, we get better control on systematic errors caused by
corrections to scaling.  Comparing the central results   
$c_{3/2} = 0.3371(28)$ and $c_{1/2} = 0.266(35)$, Ref. \cite{Banerjee18}, and
$c_{3/2} = 0.340(1)$ and $c_{1/2} = 0.23(2)$, Ref. \cite{Cuomo23}, with
our $c_{3/2} = 0.33163(9)$, Eq.~(\ref{finalC32}),  and 
$c_{1/2} = 0.2765(8)$, Eq.~(\ref{finalC32}), shows that in particular 
in Ref. \cite{Cuomo23}, systematic errors are underestimated.
The same holds for the estimates of $D_q$ as can be seen in table
\ref{smallQ}. 

In table \ref{smallQ} we compare our results with estimates of $Y_q$ 
that were obtained directly for the given value of $q$, 
by using various methods.  We see that in general the error increases 
rapidly with increasing $q$.  For $q \le 3$ our results nicely 
agree with those obtained in Ref. \cite{che19} by using the 
conformal bootstrap method. The deviation for $q=4$ between 
our estimate and that of Ref. \cite{O2corrections} is about three 
times the sum of the errors that are quoted. On the other hand,
we confirm the estimate of Ref. \cite{myCubicN2}. The estimates obtained 
by using the 5-loop $\epsilon$-expansion and the 6-loop expansion in 
three dimensions fixed are consistent with our result, but the error
is clearly larger than ours.  The same holds for Ref. \cite{Chle22}, 
where the functional renormalization group (FRG) is employed.
In the case of the Monte Carlo study of a lattice model \cite{Shao20}, 
the estimate for $q=4$ is consistent with ours, but too small 
for $q=5$ and $6$. 

\begin{table}
\caption{\sl \label{smallQ}
Estimates for the RG exponent $Y_q = 3-D_q$ obtained directly for the 
given $q \le 6$.  For $q<4$, there is an abundance
of results in the literature that we do not report here. We focus on recent
results obtained by using the CB method and Monte Carlo (MC) simulations of lattice 
models. Note that the asterisk in the case of Ref. \cite{O2corrections}
indicates that the error estimate is no rigorous bound.
For comparison, we give results obtained by using the iterative 
method in Refs. \cite{Banerjee18,Cuomo23} and here.
}
\begin{center}
\begin{tabular}{ccllllll}
\hline
Ref. & Method \textbackslash{}  $q$   & \mc{1}{c}{1}   &  \mc{1}{c}{2}  & \mc{1}{c}{3}  & \mc{1}{c}{4}  &\mc{1}{c}{5} & \mc{1}{c}{6} \\
\hline
\cite{Carmona}& $\epsilon$-exp. & 
                  &     &  & --0.114(4) & & \\
\cite{Carmona}& 3D exp.  &  &  &     & --0.103(8) & &  \\
\cite{Chle22} &FRG & &  &     & --0.111(12) & & \\
\cite{che19} & CB & 2.480912(22) & 1.76371(11)& 0.8914(3$^*$)&    &  &  \\
\cite{O2corrections}& CB& &      &            &--0.11535(73$^*$)&  &  \\
\cite{O234}   &MC &  &1.7639(11) & 0.8915(20) &--0.108(6)  &    &      \\
\cite{Shao20} & MC & &           &        &--0.114(2)  &--1.27(1)&--2.55(6) \\
\cite{myCubicN2}&MC & &         &         &--0.1118(10)&    &       \\
\cite{myClock} & MC &2.48095(4) &     &       &            &    & --2.43(6)\\
\cite{myXY2025}& MC & 2.48092(1) &             &            &    &          \\
\hline
\cite{Banerjee18}& MC & 2.484(3) & 1.762(5) & 0.884(6)   & --0.128(6) & 
 --1.265(6)    & --2.509(7) \\
\cite{Cuomo23}   & MC &  2.4801(98) & 1.761(1)  & 0.884(2) & --0.124(2) &
 --1.254(3)      & --2.486(4) \\
this work      &MC &          &1.76370(12)   & 0.89167(23) & --0.11203(34) 
  &--1.23195(47) & --2.45714(62) \\
\hline
\end{tabular} 
\end{center}
\end{table}

It had been noted before that Eq.~(\ref{CentralEQ}) fits $D_q$ down to 
surprisingly small values of $q$. This observation is further strengthened 
by our accurate estimates of $D_q$. It is remarkable that in our fits we can do
without $O(q^{-1/2})$ for $q \gtrapprox 4$. 

\section{Acknowledgement}
This work was supported by the Deutsche Forschungsgemeinschaft (DFG) under 
the grant No HA 3150/5-4.

\appendix 
\section{Ratio of partition functions}
\label{zazp_app}
The ratio of partition functions $Z_a/Z_p$ has been a cornerstone in 
our finite size scaling analysis since Ref. \cite{HaPiVi}. 
Here, $Z_a$ is the partition function
of a system with anti-periodic boundary conditions in one of the
three directions and the partition function $Z_p$ of a system with periodic
boundary conditions in all directions.  Anti-periodic boundary conditions in
$0$-direction are obtained by changing the sign of the term
$\vec{s}_x \cdot \vec{s}_y$ of the Hamiltonian for nearest neighbor pairs,
and here third nearest neighbor pairs, where $x_0=L-1$ and $y_0=0$. It turns out
that it is straight forward to determine $Z_a/Z_p$ using the worm simulation.
First we note that 
\begin{equation}
\label{signchange}
 I_k(K) = (-1)^{k} I_k(-K) \;.
\end{equation}
Let us define 
\begin{equation} 
K_{L-1,0} = \sum_{<xy>, x_0=L-1, y_0=0} k_{<xy>}  + \sum_{[xy], x_0=L-1, y_0=0} k_{[xy]}   \;.
\end{equation} 
For a given configuration $\{k\}$,  switching from periodic to anti-periodic boundary 
conditions, the product
\begin{equation}
W(\{k\})=\left[ \prod_{<xy>} I_{k_{<xy>}}(K_1) \right] \left[ \prod_{[xy]} I_{k_{[xy]}}(K_3) \right]
\end{equation}
changes sign if $K_{L-1,0}$ is odd.  We can write the partition functions, 
without insertion of head and tail, as
\begin{equation}
 Z_p = \sum_{\{k\}'} W(\{k\}) \;\;\; , \;\; Z_a=\sum_{\{k\}'} W(\{k\}) (-1)^{K_{L-1,0}} \;,
\end{equation}
where $\{k\}'$ are configurations that satisfy the constraint eq.~(\ref{const0}). 
Hence 
\begin{equation}
\frac{Z_a}{Z_p} = \frac{\sum_{\{k\}'} W(\{k\}) (-1)^{K_{L-1,0}}}
                       {\sum_{\{k\}'} W(\{k\})}  = \langle (-1)^{K_{L-1,0}} \rangle_p \;.
\end{equation}
This means that the ratio of partition functions is just the expectation value 
of $(-1)^{K_{L-1,0}}$ for periodic boundary conditions. One should note that only 
loops with non-vanishing winding number contribute to $K_{L-1,0}$. Furthermore,
$K_{L-1,0}=K_{0,1}=K_{1,2}, ...$.  
It is straight forward to compute derivatives of $Z_a/Z_p$ with respect to $K_1$ and $K_3$. 
We have implemented the first and second derivative of $Z_a/Z_p$ with respect to $K_1$
in our program. Furthermore we have implemented the internal energy and the magnetic 
susceptibility.  Using $w(x,y)=1$, without inserting charge, we performed high statistics 
simulations for smaller lattice sizes. The results are consistent with those of 
Ref. \cite{myXY2025}, which were obtained by using the cluster algorithm 
\cite{SwWa87,Wolff}. Benchmarked on the statistical error of $Z_a/Z_p$ for 
example, the performance of the worm and cluster algorithm are similar.


\begin{thebibliography}{99}

\bibitem{WiKo}
K. G. Wilson and J. Kogut,
{\sl The renormalization group and the $\epsilon$-expansion},
Phys.\ Rep.\ C {\bf 12}, 75 (1974).

\bibitem{Fisher74}
M. E. Fisher,
{\sl The renormalization group in the theory of critical behavior},
Rev.\ Mod.\ Phys.\ {\bf 46}, 597 (1974), Erratum:
Rev.\ Mod.\ Phys.\ {\bf 47}, 543 (1975).

\bibitem{Cardy}
John Cardy, {\sl Scaling and Renormalization in Statistical Physics},
Series: Cambridge Lecture Notes in Physics (No. 5)
(Cambridge University Press, Cambridge, 1996)

\bibitem{Fisher98}
M. E. Fisher,
{\sl Renormalization group theory: Its basis and formulation in statistical physics},
Rev.\ Mod.\ Phys.\ {\bf 70}, 653 (1998).

\bibitem{PeVi02}
A. Pelissetto and E. Vicari,
{\sl Critical Phenomena and Renormalization-Group Theory},
[arXiv:cond-mat/0012164],
Phys.\ Rept.\ {\bf 368}, 549 (2002).

\bibitem{PaRyVi18} 
D. Poland, S. Rychkov, and A. Vichi,
{\sl The Conformal Bootstrap: Theory, Numerical Techniques, and Applications},
[arXiv:1805.04405], Rev.\ Mod.\ Phys.\ {\bf 91}, 15002 (2019).
%

\bibitem{RySu23}
Slava Rychkov and Ning Su,
{\sl New Developments in the Numerical Conformal Bootstrap},  
[arXiv:2311.15844], Rev.\ Mod.\ Phys.\ {\bf 96}, 045004 (2024).

\bibitem{Alday}
L.F. Alday and J.M. Maldacena, {\sl Comments on operators with large spin},
[arXiv:0708.0672], J.\ High Energ.\ Phys.\  11 (2007) 019.

\bibitem{Fitz}
A. L. Fitzpatrick, J. Kaplan, D. Poland and D. Simmons-Duffin, 
{\sl The analytic bootstrap and AdS
superhorizon locality}, [arXiv:1212.3616], 
J.\ High Energ.\ Phys.\  12 (2013) 004.

\bibitem{Koma}
Z. Komargodski and A. Zhiboedov, 
{\sl Convexity and liberation at large spin}, 
[arXiv:1212.4103], 
J.\ High Energ.\ Phys.\  11 (2013) 140.

\bibitem{Hellerman2015}
S. Hellerman, D. Orlando, S. Reffert, and M. Watanabe,
{\sl On the CFT operator spectrum at large global charge}, 
[arXiv:1505.01537],
J.\ High Energ.\ Phys.\ 12  (2015) 71.

\bibitem{Monin17}
A. Monin, D. Pirtskhalava, R. Rattazzi and F.K. Seibold, 
{\sl Semiclassics, Goldstone bosons and CFT data}, 
[arXiv:1611.02912], J.\ High Energ.\ Phys.\ 06 (2017) 011.





\bibitem{Alvarez21}
Luis Alvarez-Gaume, Domenico Orlando, Susanne Reffert,
{\sl Selected topics in the large quantum number expansion},
[arXiv:2008.03308],
Phys.\ Rept.\  {\bf 933}, 1 (2021).

\bibitem{Banerjee18}
Debasish Banerjee, Shailesh Chandrasekharan, and Domenico Orlando,
{\sl Conformal Dimensions via Large Charge Expansion},
[arXiv:1707.00711], 
Phys.\ Rev.\ Lett.\ {\bf 120}, 061603 (2018).
\verb+https://doi.org/10.1103/PhysRevLett.120.061603 +

\bibitem{Prok98}
 N. V. Prokof'ev, B. V. Svistunov, S. Tupitsyn, I, ``Worm'' Algorithm in
Quantum Monte Carlo Simulations, Phys.\ Lett.\ A {\bf 238}, 253 (1998).

\bibitem{Prok01}
N. Prokof'ev, B. Svistunov, Worm Algorithms for Classical Statistical Models,
[arXiv:cond-mat/0103146],
Phys.\ Rev.\ Lett.\ {\bf 87}, 160601 (2001).

\bibitem{Cuomo23}
Gabriel Cuomo, J. M. Viana Parente Lopes, Jos\'e Matos, J\'ulio Oliveira,
and Jo\~ao Penedones,
{\sl Numerical tests of the large charge expansion},
[arXiv:2305.00499], J.\ High Energ.\ Phys.\  05 (2024) 161.

\bibitem{myXY2025}
M. Hasenbusch,
{\sl Eliminating leading and subleading corrections to scaling in the 
three-dimensional XY universality class},
[arXiv:2507.19265], Phys.\ Rev.\ B {\bf 112}, 184512 (2025).

\bibitem{Ulli_Demo_09}
Ulli Wolff,
{\sl Simulating the All-Order Strong Coupling
Expansion I: Ising Model Demo}
[arXiv:0808.3934], Nucl.\ Phys.\ B {\bf 810}, 491 (2009).

\bibitem{HaTo99}
M. Hasenbusch  and T. T\"or\"ok,
{\sl High precision Monte Carlo study of the 3D XY-universality class},
[arXiv:cond-mat/9904408], J. Phys. A: Math. Gen. {\bf 32}, 6361 (1999).

\bibitem{XY1}
M. Campostrini, M. Hasenbusch, A. Pelissetto, P. Rossi, and
E. Vicari,
{\sl Critical behavior of the three-dimensional XY universality class},
[cond-mat/0010360], Phys.\ Rev.\ B {\bf 63}, 214503 (2001).

\bibitem{XY2}
M. Campostrini, M. Hasenbusch, A. Pelissetto, and E. Vicari,
{\sl The critical exponents of the superfluid transition in He4},
[cond-mat/0605083], published as
{\sl Theoretical estimates of the critical exponents of the superfluid
transition in He4 by lattice methods}, Phys.\ Rev.\ B {\bf 74}, 144506 (2006).

\bibitem{Banerjee10}
Debasish Banerjee and Shailesh Chandrasekharan, 
{\sl Finite size effects in the presence of a chemical potential:
A study in the classical non-linear $O(2)$ sigma-model}, 
[arXiv:1001.3648], Phys.\ Rev.\ D {\bf 81} 125007, (2010). \\
\verb+https://doi.org/10.1103/PhysRevD.81.125007 +

\bibitem{Deng19}
Wanwan Xu, Yanan Sun, Jian-Ping Lv, and Youjin Deng,
{\sl High-precision Monte Carlo study of several models in the 
three-dimensional $U(1)$ universality class}, 
[arXiv:1908.10990], Phys.\ Rev.\ B {\bf 100}, 064525 (2019)
\verb+https://doi.org/10.1103/PhysRevB.100.064525+








\bibitem{myClock}
M. Hasenbusch,
{\sl Monte Carlo study of an improved clock model in three dimensions},
[arXiv:1910.05916],
Phys.\ Rev.\ B {\bf 100}, 224517 (2019).

\bibitem{Meneses19}
S. Meneses, J. Penedones, S. Rychkov, J.M. Viana Parente Lopes,
and P. Yvernaye,
{\sl A structural test for the conformal invariance of the
critical 3d Ising model},
[arXiv:1802.02319], 
J.\ High Energ.\ Phys.\  04 (2019), 115.

\bibitem{O2corrections}  
Junyu Liu, David Meltzer, David Poland, David Simmons-Duffin,
{\sl The Lorentzian inversion formula and the spectrum of the 3d O(2) CFT}
[arXiv:2007.07914], JHEP 09 (2020) 115.

\bibitem{private}
Junyu Liu and David Simmons-Duffin communicated the numerical value
$\omega_{NR}=2.02548(41)$ in private. 


\bibitem{che19}
S. M. Chester, W. Landry, J. Liu, D. Poland, D. Simmons-Duffin, N. Su,
and A. Vichi,
{\sl Carving out OPE space and precise $O(2)$ model critical exponents},
[arXiv:1912.03324], 
J.\ High Energ.\ Phys.\ 06 (2020) 142.

\bibitem{O234}
M. Hasenbusch and E. Vicari,
{\sl Anisotropic perturbations in three-dimensional O(N)-symmetric vector
 models},
[arXiv:1108.0491], Phys.\ Rev.\ B {\bf 84}, 125136 (2011).

\bibitem{Carmona}
J. M. Carmona, A. Pelissetto, and E. Vicari, 
{\sl 
$N$-component Ginzburg-Landau Hamiltonian with cubic anisotropy: 
A six-loop study}, [arXiv:cond-mat/9912115],
Phys.\ Rev.\ B {\bf 61}, 
15136 (2000).

\bibitem{Chle22}
Andrzej Chlebicki, Carlos A. S\'anchez-Villalobos, Pawel Jakubczyk, and
Nicol\'as Wschebor, 
{\sl $\mathbb{Z}_4$-symmetric perturbations to the XY model from functional
renormalization}, 
[arXiv:2204.02089], 
Phys.\ Rev.\ E {\bf 106}, 064135 (2022).

\bibitem{Shao20}
Hui Shao, Wenan Guo and Anders W. Sandvik,
{\sl Monte Carlo Renormalization Flows in the Space of Relevant and Irrelevant Operators:
Application to Three-Dimensional Clock Models},
[arXiv:1905.13640], Phys.\ Rev.\ Lett.\ {\bf 124}, 080602 (2020).


\bibitem{myCubicN2}
M. Hasenbusch, 
{\sl Monte Carlo study of the $O(2)$-invariant $\phi^4$ theory with
a cubic perturbation in three dimensions}, [arXiv:2510.19473].

\bibitem{HaPiVi}
M. Hasenbusch, K. Pinn, and S. Vinti,
{\sl Critical Exponents of the 3D Ising Universality Class From Finite Size
Scaling With Standard and Improved Actions},
[arXiv:hep-lat/9806012], Phys.\ Rev.\ B {\bf 59}, 11471 (1999).

\bibitem{SwWa87}
Robert H. Swendsen and Jian-Sheng Wang,
{\sl Nonuniversal critical dynamics in Monte Carlo simulations}, 
Phys.\ Rev.\ Lett.\ {\bf 58}, 86 (1987). 

\bibitem{Wolff}
U. Wolff,
{\sl Collective Monte Carlo Updating for Spin Systems},
Phys.\ Rev.\ Lett.\ {\bf 62}, 361 (1989).

\end{thebibliography}
\end{document}